\documentclass[runningheads]{llncs}

\usepackage[T1]{fontenc}
\usepackage[utf8]{inputenc}
\usepackage{graphicx}
\usepackage{amsmath}
\usepackage{amssymb}
\usepackage{booktabs}
\usepackage{algorithm}
\usepackage{algpseudocode}
\usepackage{listings}
\usepackage{xcolor}
\usepackage{url}
\usepackage[hidelinks]{hyperref}

\newcommand{\safefig}[2][width=\linewidth]{%
  \IfFileExists{#2}{\includegraphics[#1]{#2}}%
  {\fbox{\parbox[c][3.2cm][c]{0.92\linewidth}{\centering
     \textbf{[figure file not found]}\\[2pt]\texttt{\detokenize{#2}}}}}%
}

\newcommand{\diffF}{\texttt{diff\_F1}}

\lstdefinestyle{csnip}{
  language=C,
  basicstyle=\ttfamily\scriptsize,
  commentstyle=\itshape\color{black!60},
  keywordstyle=\bfseries,
  showstringspaces=false,
  columns=fullflexible,
  breaklines=true,
  frame=none,
  xleftmargin=0.5em
}
\begin{document}

\title{Metrics Failure in LLM-Based Code Vulnerability Repair: An Empirical Study and a Change-Aware Screen}
\titlerunning{Metrics Failure in LLM Code Vulnerability Repair}

\author{Om Nepal \and Sushant Aryal \and Oluseyi Olukola \and Nick Rahimi}
\authorrunning{O. Nepal et al.}

\institute{School of Computing Sciences and Computer Engineering,\\
The University of Southern Mississippi, Hattiesburg, MS 39406, USA\\
\email{\{om.nepal,sushant.aryal,oluseyi.olukola,nick.rahimi\}@usm.edu}}

\maketitle

% =====================================================================
\begin{abstract}
Large language models (LLMs) are increasingly applied to the automated repair of C/C++ security vulnerabilities, and compile rate is a commonly reported proxy for progress: whether the generated patch compiles. We argue that compile rate is a scientifically unreliable metric for single-function vulnerability repair, and we support this with five controlled experiments over 203 vulnerable functions from Big-Vul, three open-source code LLMs (350M to 6.7B parameters), and three prompting strategies. Compile rate (i) barely responds to an intervention that substantially improves the generated code; (ii) is dominated by evaluation-harness and dataset artifacts rather than model quality, with about 64\% of compile failures not attributable to the model, a share that is nearly invariant across models; (iii) shifts by 1.8 to 2.7 times on identical patches under a single compiler-standard flag, with zero regressions; (iv) ranks the three models in the opposite order to reference-similarity metrics; and (v) rewards non-repairs when used as an optimization target, since a compiler-feedback loop raises compile rate while similarity to the human fix falls, with manual inspection finding deletion- and placeholder-style non-repairs among the newly compiling outputs. The natural fallback, whole-function CodeBLEU, also fails: an unchanged copy of the vulnerable input outscores every model. We also examine \diffF, a change-aware screen that scores only the edited region. It gives exactly zero credit to a no-op and near-zero credit to some, though not all, of the deletion-based gaming patches we observed, while still crediting genuine partial edits, so it may serve as a cheap screen before deeper, execution-based analysis. It is not a repair-quality metric, and we report where it falls short. Our findings argue for change-aware, execution-grounded evaluation of LLM-based vulnerability repair.

\end{abstract}

% =====================================================================
\section{Introduction}
\label{sec:intro}

Memory-unsafe languages remain the backbone of critical software: operating systems, browsers, language runtimes, and embedded firmware are overwhelmingly written in C and C++, and they continue to be a dominant source of exploitable security vulnerabilities, from buffer overflows and out-of-bounds accesses to integer errors and improper input validation \cite{fan2020bigvul}. Patching such vulnerabilities by hand is slow and expensive, which has motivated a large and growing body of work on automated program repair and, most recently, on applying LLMs trained on source code to generate candidate fixes directly from vulnerable functions \cite{basic2024slr,nong2025appatch}. Open-source code LLMs such as CodeGen \cite{nijkamp2023codegen} and DeepSeek-Coder \cite{guo2024deepseek} have made this line of work broadly reproducible, and a steady stream of papers now reports that LLMs can repair some fraction of real-world vulnerabilities.

Any such claim of progress rests on a measurement. Whether a generated patch is good should mean that it removes the vulnerability without breaking the intended behavior of the program, a property that can be checked directly only with executable tests or proof-of-concept exploits. For many function-level, CVE-derived datasets, executable tests or exploit oracles are not supplied with each example, so the community falls back on cheap, automatic proxies. One of the most commonly reported is compile rate: the fraction of generated patches that build, equivalent to pass@1 under uniform sampling \cite{chen2021codex}. A frequent secondary proxy is textual similarity to the fix written by the developer, typically CodeBLEU \cite{ren2020codebleu}. These proxies are appealing precisely because they are automatic and require no per-example test harness. Yet their validity as measures of repair quality has received little scrutiny, especially in the single-function setting that dominates this literature, where a function is evaluated in isolation from its surrounding project.

This paper subjects that assumption to controlled scrutiny and finds it wanting. Using 203 vulnerable functions from Big-Vul \cite{fan2020bigvul}, three open-source code LLMs spanning 350M to 6.7B parameters, and three prompting strategies, we design five controlled experiments, each isolating a distinct way in which an evaluation metric can be unreliable: insensitivity to genuine improvement, confounding by the measurement setup, dependence on the toolchain configuration, disagreement with an independent metric, and gameability under optimization. Compile rate exhibits all five failure modes. Worse, the commonly used reference-similarity fallback, whole-function CodeBLEU, is itself gameable: an unmodified copy of the vulnerable input scores higher than every model we test, because most of a correct fix is unchanged surrounding context. Motivated by these findings, we examine \diffF, an exploratory change-aware screen, not a repair-quality metric, which considers only the edited region. It assigns zero credit to no-ops and near-zero credit to some code-deletion patches induced by optimizing for compilation, while still crediting genuine partial repairs. It does not do so for every such patch, and we report where it falls short.

\subsection*{Contributions}

% llncs.cls sets \labelitemi to a bold dash. Override it inside this
% group only, so the contributions list uses round bullets while any
% other list in the paper keeps the LNCS default.
\begingroup
\renewcommand{\labelitemi}{\textbullet}
\begin{itemize}
  \item \textbf{An evaluation critique of compile rate.} Through five independent and mutually reinforcing controlled experiments (Section~\ref{sec:results}), we show that compile rate is an unreliable metric for single-function LLM-based vulnerability repair. It is not merely noisy, but systematically misleading in five distinct ways.
  \item \textbf{A measurement-setup explanation of compile failures.} We show that the majority, approximately 64\%, of compile failures in this setting are not caused by the model writing bad code, but by evaluating a function in isolation (missing project context) and by a specific, under-documented Big-Vul preprocessing artifact. This share is nearly invariant across three very differently sized models (Section~\ref{subsec:failures}).
  \item \textbf{A demonstrated failure of the similarity fallback.} We show that whole-function CodeBLEU rewards preserved context rather than repair: a no-op copy of the vulnerable function outscores every model (Section~\ref{subsec:disagreement}).
  \item \textbf{A change-aware screen.} We examine \diffF, a change-aware edit-overlap score, and show that it gives zero credit to no-ops and near-zero credit to some observed gaming patches while assigning nonzero credit to model edits that overlap the developer's edit (Section~\ref{subsec:difff1}). Its credit for other deletion-style patches can be considerably higher (Section~\ref{sec:limitations}), so we frame it as a screen before deeper analysis, not as a measure of repair quality.
  \item \textbf{A benchmark-scale repair comparison.} As a secondary contribution, we report a three-model, three-strategy repair comparison at a 203-function scale with bootstrap confidence intervals, substantially larger than the tens of examples typical of this line of work (Section~\ref{subsec:overview}).
\end{itemize}
\endgroup

The rest of this paper is organized as follows. Section~\ref{sec:related} surveys related work on vulnerability datasets, LLM-based repair, and code-generation evaluation metrics. Section~\ref{sec:method} describes the dataset, a data-quality artifact we identify, the models and prompting strategies, the isolated-function compile harness, the metrics, and the design of the five controlled experiments. Section~\ref{sec:results} presents the results, including the change-aware screen. Section~\ref{sec:discussion} discusses the implications. Section~\ref{sec:limitations} states the limitations and future work, and Section~\ref{sec:conclusion} concludes.

% =====================================================================
\section{Related Work}
\label{sec:related}

\paragraph{Vulnerability datasets and their quality.}
Datasets of real-world vulnerabilities underpin empirical work on automated repair. Big-Vul \cite{fan2020bigvul} is among the most widely used: it links over 3{,}700 C/C++ vulnerabilities from 348 open-source projects to their fixing commits and CVE metadata, and we draw our evaluation functions from it. A recurring concern with such Git-mined datasets is label noise and duplication. PrimeVul \cite{ding2025primevul} shows that on cleaner, de-duplicated data the apparent vulnerability-detection ability of code LLMs drops sharply, implying that numbers derived from Big-Vul should be read cautiously; a recent systematic review \cite{basic2024slr} surveys LLMs across detection, generation, and patching and catalogues these data-quality issues. Big-Vul is also not the only dataset built by this recipe: CVEfixes \cite{bhandari2021cvefixes} mines CVE records and their fixing commits at a larger scale, so the preprocessing decisions taken inside such pipelines propagate into every study built on top of them, a point we return to in Section~\ref{subsec:artifact}. Our critique is largely orthogonal to dataset quality because our strongest result, compiler-flag sensitivity on identical patches (Section~\ref{subsec:flags}), holds regardless of label correctness, and we additionally document a specific, under-reported Big-Vul preprocessing artifact (Section~\ref{subsec:artifact}).

\paragraph{LLM-based program and vulnerability repair.}
A rapidly growing line of work applies LLMs to program repair and, specifically, to vulnerability patching. Open code LLMs such as CodeGen \cite{nijkamp2023codegen} and DeepSeek-Coder \cite{guo2024deepseek} are common backbones, and recent systems such as APPATCH \cite{nong2025appatch} combine vulnerability-semantics reasoning with adaptive prompting to generate patches without fine-tuning or exploit evidence. Earlier learning-based systems fine-tune sequence-to-sequence models for the same task: VulRepair \cite{fu2022vulrepair} adapts a pre-trained T5 model to vulnerability repair and reports its headline result as a perfect prediction rate, the share of functions whose generated fix matches the fix written by the developer exactly, which shows how directly this literature equates success with reproducing the reference output as a whole. Zhang et al. \cite{zhang2024survey} survey learning-based program repair more broadly and document the same dependence on automatic, reference-based or test-based patch assessment. This literature focuses on improving repair; we instead ask whether the metrics used to judge such repairs are trustworthy. Our study deliberately uses the same base models and prompting strategies (zero-shot, few-shot, and chain-of-thought) as prior work, ensuring that our critique targets realistic evaluation conditions rather than a strawman.

\paragraph{Evaluation metrics for code generation.}
Two proxy metrics dominate code-generation evaluation. pass@k, introduced with Codex and HumanEval \cite{chen2021codex}, estimates the probability that at least one of $k$ samples passes a set of tests; in the frequent absence of executable tests for real CVEs, it collapses to whether the code compiles, since pass@1 equals compile rate under uniform sampling. CodeBLEU \cite{ren2020codebleu} augments BLEU with abstract-syntax-tree and data-flow matching to compare generated code against a reference fix. The limits of such reference-based scores are themselves documented: Evtikhiev et al. \cite{evtikhiev2023bleu} find that BLEU-style metrics, CodeBLEU among them, agree only weakly with human assessments of generated code and can reorder systems relative to human preference. Where a per-example execution oracle can be built, the field prefers it, as in SWE-bench \cite{jimenez2024swebench}, which accepts a patch only if the test suite of the project passes; building such a harness for each of thousands of CVE-derived functions is not currently practical, which is why the cheap proxies persist in vulnerability repair. Both pass@k and CodeBLEU were designed for whole-program or whole-function generation; neither was designed to isolate the small edit that a vulnerability fix typically represents. We show that this mismatch is not benign.

\paragraph{Critiques of secure-code and repair evaluation.}
A small but growing body of work questions how secure-code and repair systems are evaluated. Most closely related, Dai et al. \cite{dai2025rethinking} argue that secure-code-generation evaluations conflate security with functionality and over-rely on a single analyzer, and find that state-of-the-art methods often achieve security by removing functionality, that is, by deleting code. We independently observe the same code-deletion behavior when compile rate is used as an optimization target (Section~\ref{subsec:gaming}), from a different angle (compilation rather than static security analysis) and on a real CVE dataset. Where Dai et al. propose an execution-grounded combined metric, we examine a possible complementary, lightweight change-aware score that requires no tests. Other work explores an LLM-as-a-judge for patch evaluation \cite{shi2025judge}; such judges, like execution-based oracles, are heavier-weight and complementary to the cheap textual metric we study. The same reliance on a single automatic proxy is visible in secure-code generation, where SVEN \cite{he2023sven} hardens code LLMs and quantifies the outcome as a security rate computed by a static analyzer over a fixed prompt set. To our knowledge, no prior work isolates and quantifies the specific ways in which compile rate fails as a metric for single-function vulnerability repair.

\paragraph{Summary of the gap.}
Table~\ref{tab:relatedwork} summarizes this gap. What the prior work above has in common is not the oracle it uses, which ranges from unit tests to static analysis to an LLM judge, but the unit it scores: the generated output taken as a whole. None of it scores the edit that a vulnerability fix actually consists of. That is the column our work explores.

\begin{table}[t]
\centering
\caption{How repair and code-generation work is evaluated. Whole-output evaluation means the generated function or program is judged as a whole, by whatever oracle: unit tests for pass@k \cite{chen2021codex} and SWE-bench \cite{jimenez2024swebench}, exact match to the fix written by the developer for VulRepair \cite{fu2022vulrepair}, reference similarity for CodeBLEU \cite{ren2020codebleu}, a static analyzer plus execution for Dai et al. \cite{dai2025rethinking}, and an LLM judge for \cite{shi2025judge}. Change-aware evaluation means the score is computed on the edit rather than on the whole output. PrimeVul is a detection benchmark, so whole-output repair evaluation does not apply to it.}
\label{tab:relatedwork}
\begin{tabular}{@{}llcc@{}}
\toprule
Paper & Dataset & Whole-output eval. & Change-aware eval. \\
\midrule
Codex \cite{chen2021codex}        & HumanEval               & yes & no \\
CodeBLEU \cite{ren2020codebleu}   & code-gen. benchmarks    & yes & no \\
SWE-bench \cite{jimenez2024swebench} & GitHub issues        & yes & no \\
VulRepair \cite{fu2022vulrepair}  & Big-Vul, CVEfixes       & yes & no \\
APPATCH \cite{nong2025appatch}    & zero-day, ExtractFix    & yes & no \\
Dai et al. \cite{dai2025rethinking} & secure-gen. sets      & yes & no \\
LLM-judge \cite{shi2025judge}     & generated patches       & yes & no \\
PrimeVul \cite{ding2025primevul}  & PrimeVul (detection)    & n/a & no \\
\midrule
This work                         & Big-Vul (203 fns)       & yes & yes (\diffF) \\
\bottomrule
\end{tabular}
\end{table}

% =====================================================================
\section{Methodology}
\label{sec:method}

\begin{figure}[t]
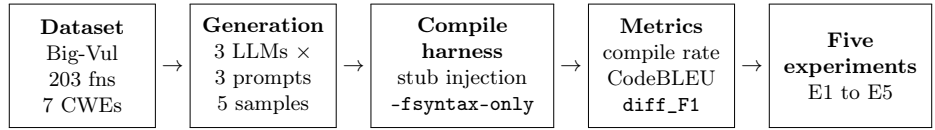

\centering
\setlength{\fboxsep}{4pt}
\footnotesize
% The five stages must sit on ONE line. Their natural width (boxes + \fboxsep +
% rules + arrows) slightly exceeds \linewidth, which previously pushed the fifth
% box onto a second row; \resizebox guarantees a single row.
\resizebox{\linewidth}{!}{%
\fbox{\parbox[c][1.55cm][c]{0.155\linewidth}{\centering \textbf{Dataset}\\ Big-Vul\\ 203 fns\\ 7 CWEs}}%
$\;\rightarrow\;$%
\fbox{\parbox[c][1.55cm][c]{0.155\linewidth}{\centering \textbf{Generation}\\ 3 LLMs $\times$\\ 3 prompts\\ 5 samples}}%
$\;\rightarrow\;$%
\fbox{\parbox[c][1.55cm][c]{0.20\linewidth}{\centering \textbf{Compile}\\ \textbf{harness}\\ stub injection\\ \texttt{-fsyntax-only}}}%
$\;\rightarrow\;$%
\fbox{\parbox[c][1.55cm][c]{0.155\linewidth}{\centering \textbf{Metrics}\\ compile rate\\ CodeBLEU\\ \diffF}}%
$\;\rightarrow\;$%
\fbox{\parbox[c][1.55cm][c]{0.175\linewidth}{\centering \textbf{Five}\\ \textbf{experiments}\\ E1 to E5}}%
}
\caption{Overview of the evaluation pipeline. Each vulnerable function from the Big-Vul test set (Section~\ref{subsec:dataset}) is patched by three LLMs under three prompting strategies with five samples per cell; every patch is syntax-checked by the stub-injection harness of Section~\ref{subsec:harness}, scored by the metrics of Section~\ref{subsec:metrics}, and the scored patches feed the five controlled experiments of Section~\ref{subsec:design}.}
\label{fig:pipeline}
\end{figure}

This section describes the dataset and a data-quality artifact we identify, the models and prompting strategies we evaluate, the generation configuration, the isolated-function compile harness that underlies our compile-rate measurements, the metrics we report, and the design of the five controlled experiments. 
Figure~\ref{fig:pipeline} gives an overview. Except where noted, the setup mirrors common practice in LLM-based vulnerability repair, so that our critique applies to how progress is actually measured in this line of work.

\subsection{Dataset}
\label{subsec:dataset}

We evaluate on Big-Vul \cite{fan2020bigvul}, a widely used dataset of real C/C++ vulnerabilities mined from public GitHub projects. Each entry links a pre-fix function (\texttt{func\_before}) to the post-fix function written by the developer (\texttt{func\_after}) together with the associated CVE and CWE metadata. We treat \texttt{func\_after} as the ground-truth human fix throughout, and the CWE label as the only side information provided to the model. Big-Vul is one of several datasets assembled by mining CVE records and their fixing commits; CVEfixes \cite{bhandari2021cvefixes} later applied the same recipe to over 5{,}000 CVE records across more than 1{,}700 open-source projects. We evaluate on Big-Vul because it remains the de facto standard in LLM-based C/C++ repair studies, which makes it the appropriate target for a critique of how those studies are measured.

\paragraph{Test set.}
We draw a stratified test set of 203 vulnerable functions spanning seven CWE categories, with 29 functions per category (Table~\ref{tab:testset}). All seven are among the most frequent categories in Big-Vul, and together they cover the memory-handling and input-handling weaknesses that dominate C/C++ CVEs. We additionally hold out a disjoint few-shot pool of three examples per category, which are never included in the test set, to supply the exemplars used by the few-shot prompting strategy (Section~\ref{subsec:models}). This is a focused evaluation on these seven classes rather than a survey of every category in the dataset, and we make no claim about categories outside it.

\paragraph{A stratified subset for paired ablations.}
Some of our controlled experiments, such as the generation-budget comparison and the compiler-standard sweep, require repeatedly re-generating or re-validating patches under multiple configurations. To bound compute cost for these, we use a fixed, stratified 70-function subset of the 203 (the first ten functions per category under the same seeded draw). This is a strict, non-cherry-picked subset: filtering the full 203-function results down to these 70 function identifiers reproduces the numbers of the subset exactly, to the decimal. Throughout, population-level estimates (headline compile rate, CodeBLEU, and failure-category shares) are reported on the full 203 functions. Paired, within-subject ablations that hold the function set fixed and vary a single factor are reported on the 70-function subset where compute required it, and on the full 203 functions otherwise. We state which sample size applies to each result.

\begin{table}[t]
\centering
\caption{Composition of the 203-function Big-Vul test set: seven CWE categories, 29 functions each. A disjoint pool of three examples per category is held out for few-shot prompting.}
\label{tab:testset}
\begin{tabular}{@{}llr@{}}
\toprule
CWE & Description & \# Functions \\
\midrule
CWE-119 & Improper restriction of operations within memory bounds & 29 \\
CWE-125 & Out-of-bounds read & 29 \\
CWE-189 & Numeric errors & 29 \\
CWE-20  & Improper input validation & 29 \\
CWE-200 & Exposure of sensitive information & 29 \\
CWE-264 & Permissions, privileges, and access control & 29 \\
CWE-399 & Resource management errors & 29 \\
\midrule
\textbf{Total} & & \textbf{203} \\
\bottomrule
\end{tabular}
\end{table}

\subsection{A Big-Vul Preprocessing Artifact}
\label{subsec:artifact}

While analyzing why patches fail to compile (Section~\ref{subsec:failures}), we identified a data-quality issue in Big-Vul worth reporting on its own. One distinct class of failures was functions that do not compile because they have no visible return type. We found that in 96 to 100\% of those cases, the preprocessing of Big-Vul itself had already stripped the return type from the stored \texttt{func\_before}: the model is faithfully reproducing a function signature that was malformed in the input. This class accounts for roughly 15 to 19\% of all compile failures (Section~\ref{subsec:failures}). It is neither a model error nor an artifact of our compile harness; it is a property of the dataset. Consequently, a non-trivial share of the failures attributed to models in compile-rate studies based on Big-Vul is, in fact, inherited from dataset preprocessing, representing a concrete instance of the more general problem this paper documents.

\subsection{Models, Prompting, and Generation}
\label{subsec:models}

We evaluate three open-source code LLMs spanning a nearly twentyfold range in parameter count (Table~\ref{tab:models}): CodeGen-350M-multi \cite{nijkamp2023codegen} and the 1.3B and 6.7B base variants of DeepSeek-Coder \cite{guo2024deepseek}. All three are base (non-instruction-tuned) models, the setting in which compile rate is most commonly reported. For one robustness experiment, the compiler-feedback loop of Section~\ref{subsec:gaming}, we additionally use the instruction-tuned DeepSeek-Coder-1.3B.

\begin{table}[t]
\centering
\caption{The three code LLMs in the primary comparison. All are base models; an instruction-tuned DeepSeek-Coder-1.3B is used only for a robustness experiment.}
\label{tab:models}
\begin{tabular}{@{}lllr@{}}
\toprule
Model & Hugging Face identifier & Params & Context \\
\midrule
CodeGen-350M-multi   & \texttt{Salesforce/codegen-350M-multi}      & 350M & 2{,}048 \\
DeepSeek-Coder-1.3B  & \texttt{deepseek-ai/deepseek-coder-1.3b-base} & 1.3B & 16{,}384 \\
DeepSeek-Coder-6.7B  & \texttt{deepseek-ai/deepseek-coder-6.7b-base} & 6.7B & 16{,}384 \\
\bottomrule
\end{tabular}
\end{table}

We compare three prompting strategies drawn from standard practice: zero-shot (the vulnerable function and its CWE label, with an instruction to return a fixed function), few-shot (the same, preceded by two held-out repair examples of the same CWE drawn from the few-shot pool), and chain-of-thought (the zero-shot prompt augmented with a request to reason about the vulnerability before emitting the fix). The three exact prompt templates are released verbatim with the replication package.

For every (model, strategy, function) cell we draw five independent samples with nucleus sampling at temperature 0.8 and a budget of 512 new tokens. Because CodeGen-350M has only a 2{,}048-token context, its prompt is capped at 1{,}536 tokens to leave room for generation; the 16{,}384-token context of the DeepSeek models imposes no such constraint. This yields $203 \times 3 \times 5 = 3{,}045$ patches per model in the primary comparison.

\subsection{Isolated-Function Compile Harness}
\label{subsec:harness}

Measuring whether a generated function compiles is not straightforward when the function is evaluated in isolation from its project. Real vulnerable functions reference project-specific types, macros, and helper functions that are absent from a single-function snippet, so a naive \texttt{gcc -fsyntax-only} \cite{gcc} would reject almost everything for reasons unrelated to the edit made by the model. To measure compilability fairly, we built an iterative stub-injection harness (Algorithm~\ref{alg:stub}) that synthesizes minimal declarations for the identifiers a snippet references but does not define, then syntax-checks the augmented translation unit. Figure~\ref{fig:stubexample} shows a worked example.

The harness first prepends standard headers and a small set of common typedefs (for example \texttt{u32} and \texttt{BOOL}). It compiles the candidate with \texttt{gcc/g++} \texttt{-fsyntax-only}, parses the diagnostics for missing-symbol errors (undeclared identifiers, unknown types, implicit declarations), and synthesizes a stub for each: unknown functions become variadic declarations (\texttt{int f(...);}), unknown types become aliases of a pointer to a shared generic structure whose fields grow with each member-access error, incomplete structures receive a padded body, and unknown value identifiers become plain \texttt{int} variables. Candidates that contain C++ markers are compiled with \texttt{g++}, and C candidates that still fail are retried once as C++ when they contain the scope operator. The harness reconciles the new stubs against existing declarations to avoid redeclaration collisions, prepends them, and repeats, up to six iterations, until the unit compiles or no new stubs can be added.

\begin{algorithm}[t]
\caption{Iterative stub-injection compile check}
\label{alg:stub}
\begin{algorithmic}[1]
\Require candidate function $s$; iteration cap $M$
\Ensure \textbf{true} if $s$ syntax-checks, else \textbf{false}
\State $u \gets \textsc{Wrap}(s)$ \Comment{standard headers and common typedefs around $s$}
\For{$i \gets 1$ \textbf{to} $M$}
  \State $(ok, E) \gets \textsc{Compile}(u)$ \Comment{\texttt{gcc/g++} \texttt{-fsyntax-only}}
  \If{$ok$} \State \Return \textbf{true} \EndIf
  \State $D \gets \textsc{MissingSymbols}(E)$ \Comment{undeclared ids, unknown types}
  \If{$D = \emptyset$} \State \Return \textbf{false} \Comment{failure is not a missing-symbol error} \EndIf
  \State $S \gets \textsc{Synthesize}(D)$ \Comment{variadic stubs, struct-pointer typedefs, int variables}
  \State $S \gets \textsc{Reconcile}(S, u)$ \Comment{drop collisions and redeclarations}
  \If{$S = \emptyset$} \State \Return \textbf{false} \Comment{no progress possible} \EndIf
  \State $u \gets \textsc{Prepend}(S, u)$
\EndFor
\State \Return \textbf{false}
\end{algorithmic}
\end{algorithm}

\begin{figure}[t]
\begin{lstlisting}
/* (a) Function under test, extracted from its project (isolated) */
int handle(cJSON *item, int flags) {
    if (!validate(item))          /* validate: undefined     */
        return ERR_INVALID;       /* ERR_INVALID: undefined  */
    return commit(item, flags);   /* commit: undefined       */
}

/* (b) Declarations the harness synthesizes and prepends before (a) */
struct __field_holder { int _pad; }; /* shared generic structure     */
typedef struct __field_holder *cJSON; /* unknown type -> struct pointer */
int validate(...);                /* unknown func  -> variadic stub  */
int commit(...);                  /* unknown func  -> variadic stub  */
int ERR_INVALID;                  /* unknown ident -> int variable   */
\end{lstlisting}
\caption{Worked example of stub injection. The isolated function (a) references identifiers with no definition in scope; the harness synthesizes minimal stubs (b), after which \texttt{gcc -fsyntax-only} accepts the unit.}
\label{fig:stubexample}
\end{figure}

\paragraph{Toolchain.}
All compile checks in this paper use GCC 15.2.0 (MSYS2 distribution) targeting \texttt{x86\_64-w64-mingw32} on Windows 11, invoked as \texttt{gcc} \texttt{-fsyntax-only} for C and \texttt{g++} \texttt{-fsyntax-only} for C++, with no explicit \texttt{-std} flag unless stated otherwise. The default language standard of this compiler version is C23, and \texttt{\_\_STDC\_VERSION\_\_} is \texttt{202311L}. We state this explicitly because, as Section~\ref{subsec:flags} shows, the measured compile rate is not interpretable without it.

Two design choices are load-bearing and were the subject of harness fixes during development. First, function stubs must be variadic: under modern C a declaration such as \texttt{int f()} means no arguments (equivalent to \texttt{(void)}), so a stubbed call with arguments would spuriously fail with a too-many-arguments diagnostic, whereas \texttt{(...)} accepts any argument list. Second, name reconciliation must drop a synthesized stub when the identifier is in fact declared elsewhere in the snippet, or the redeclaration itself becomes a compile error. As Sections~\ref{subsec:failures} and \ref{subsec:flags} show, the outcome of this measurement depends heavily on such harness and toolchain choices, which is precisely part of our argument.

\subsection{Evaluation Metrics}
\label{subsec:metrics}

We report three families of metrics.

\paragraph{Compile rate and pass@k.}
The compile rate is the fraction of generated patches that syntax-check under the harness of Section~\ref{subsec:harness}. Following \cite{chen2021codex}, pass@k estimates the probability that at least one of $k$ samples is acceptable, using the unbiased estimator $\text{pass@}k = 1 - \binom{n-c}{k} / \binom{n}{k}$ over $n$ samples of which $c$ compile. In our compile-based evaluation, a sample is marked successful if it passes the syntax-checking harness. Under this binary acceptance criterion and uniform sampling, the aggregate pass@1 estimator equals aggregate compile rate. We therefore treat them as one signal and do not present them as independent findings.

\paragraph{Reference similarity.}
We compute CodeBLEU \cite{ren2020codebleu} between each patch and the human fix \texttt{func\_after}, over the whole function. This is the natural fallback when execution-based checking is unavailable.

\paragraph{Change-aware similarity (\diffF).}
Because whole-function similarity is dominated by unchanged context (Section~\ref{subsec:disagreement}), we examine \diffF, which compares only the edits. It is computed on a line-level diff produced by the \texttt{SequenceMatcher} routine of Python's \texttt{difflib} module, applied to the lines of each function. For the developer, let $A_h$ and $R_h$ be the sequences of lines added and removed going from \texttt{func\_before} to \texttt{func\_after}, in diff order and with duplicates preserved; for the model, let $A_m$ and $R_m$ be the corresponding sequences going from \texttt{func\_before} to the patch. Each sequence is tokenized into a multiset of tokens by taking maximal runs of word characters as single tokens and every other non-whitespace character as its own token, without case folding and without stripping comments. Let $\mathrm{F}_1(X, Y)$ be the token-level F1 between the token multisets of a reference sequence of lines $X$ and a predicted sequence $Y$, defined as zero when either multiset is empty or when they share no token. With $S \subseteq \{A, R\}$ the sides on which the developer changed at least one line,

\begin{equation}
\label{eq:difff1}
\diffF = \frac{1}{|S|} \sum_{s \in S} \mathrm{F}_1\!\left(s_h,\, s_m\right).
\end{equation}

A patch that changes nothing has $A_m = R_m = \emptyset$ and therefore scores exactly 0, regardless of how much unchanged context it preserves; a patch whose edit exactly matches that of the developer scores 1. Every function in our test set has at least one line changed by the developer, so none is excluded. We treat \diffF{} as a change-aware screen for no-op and some deletion-style patches rather than as a measure of repair quality, and Section~\ref{subsec:difff1} and Section~\ref{sec:limitations} examine where it does and does not behave that way.

\paragraph{A note on static analysis.}
We also ran the Semgrep static analyzer \cite{semgrep}, but found it degenerate in this setting. Of the 9{,}135 generated patches, four consist only of a stray closing brace or a bare code fence, and are empty or a single character once the fence markers are stripped, so they cannot form a translation unit and the pipeline records them as degenerate output without invoking the analyzer. Across the remaining 9{,}131 patches, Semgrep returned no findings at all, reporting every one of them as vulnerability-free, including all 372 patches that compile. This is expected behaviour rather than a defect of the tool: it reports code as vulnerability-free when it cannot parse that code, which covers the large majority of our patches, and its default ruleset covers few of our target CWEs. We therefore base no claim in this paper on static-analysis output and do not report it as a result.

\subsection{Experimental Design}
\label{subsec:design}

We designed five controlled experiments, each isolating one way in which an evaluation metric can be unreliable. Table~\ref{tab:design} maps each experiment to the failure mode it probes, the factor varied, and the sample size used.

\begin{table}[t]
\centering
\caption{The five controlled experiments. Each isolates a distinct failure mode of an evaluation metric.}
\label{tab:design}
\begin{tabular}{@{}llll@{}}
\toprule
& Failure mode probed & Factor varied & $n$ \\
\midrule
E1 & Insensitivity to improvement & generation budget (256 vs 512) & 70 \\
E2 & Confounding by measurement setup & none (failure attribution) & 203 \\
E3 & Toolchain dependence & \texttt{-std} flag, patches fixed & 70, 203 \\
E4 & Disagreement with an independent metric & none (ranking comparison) & 203 \\
E5 & Gameability under optimization & compiler-feedback rounds & 203 \\
\bottomrule
\end{tabular}
\end{table}

Headline metrics report bootstrap 95\% confidence intervals (CIs) from function-level resampling with 2{,}000 replicates. E1, E3, and E5 are paired, within-subject designs: the function set is held fixed and exactly one factor is varied, so the comparison of interest is the change in each metric rather than its absolute level.

% =====================================================================
\section{Results}
\label{sec:results}

Unless stated otherwise, results are on the full 203-function test set; some paired ablations use the 70-function subset (Section~\ref{subsec:dataset}), and we state the sample size in each case.

\subsection{Three-Model, Three-Strategy Comparison}
\label{subsec:overview}

Table~\ref{tab:headline} reports each compile rate, pass@5, and CodeBLEU on the full test set, with bootstrap 95\% CIs. The two metrics tell opposite stories: the smallest model, CodeGen-350M, has the highest compile rate (5.45\%) but the lowest CodeBLEU (0.463), while the largest, DeepSeek-6.7B, has the lowest compile rate (3.32\%) but the highest CodeBLEU (0.675). Compile rate and reference similarity thus rank the three models in reverse order. We examine this disagreement in Section~\ref{subsec:disagreement}.

A second observation previews a discriminative-power problem. Compile rate places CodeGen highest, but its interval overlaps those of both DeepSeek models, and the two DeepSeek intervals, $[1.38, 6.04]$ and $[1.22, 5.88]$, almost entirely overlap each other, so compile rate separates none of the three models on unpaired intervals. CodeBLEU separates CodeGen from both DeepSeek models with non-overlapping intervals and orders the two DeepSeek models consistently, although their intervals ($[0.637, 0.685]$ and $[0.650, 0.699]$) partially overlap under unpaired resampling. Compile rate therefore has the lower resolving power here, but neither metric cleanly separates the two DeepSeek models on unpaired intervals.

\begin{table}[t]
\centering
\caption{Headline metrics on the full 203-function test set, with bootstrap 95\% confidence intervals. Compile rate ranks the models CodeGen $>$ 1.3B $>$ 6.7B; CodeBLEU ranks them in the opposite order. The two DeepSeek models are statistically indistinguishable on compile rate and only weakly separated on CodeBLEU.}
\label{tab:headline}
\begin{tabular}{@{}lccc@{}}
\toprule
Model & Compile \% [95\% CI] & Pass@5 \% & CodeBLEU [95\% CI] \\
\midrule
CodeGen-350M  & 5.45 [3.51, 7.75] & 14.1 & 0.463 [0.441, 0.484] \\
DeepSeek-1.3B & 3.45 [1.38, 6.04] & 4.8  & 0.662 [0.637, 0.685] \\
DeepSeek-6.7B & 3.32 [1.22, 5.88] & 4.8  & 0.675 [0.650, 0.699] \\
\bottomrule
\end{tabular}
\end{table}

\paragraph{Prompting strategy.}
Table~\ref{tab:strategy} breaks compile rate down by prompting strategy on the full test set. The effect of prompting is small: within any model the spread across the three strategies is at most 2.0 percentage points. Zero-shot prompting attains the highest compile rate for all three models, so neither in-context examples nor chain-of-thought reasoning improves this metric, and the relative ordering of few-shot and chain-of-thought is not consistent across models. This is itself consistent with the central argument of the paper, since compile rate responds weakly to interventions that plausibly change output quality.

\begin{table}[t]
\centering
\caption{Compile rate (\%) by model and prompting strategy on the full 203-function test set under the default toolchain. Each cell covers 1{,}015 patches (203 functions $\times$ 5 samples), and each row averages to that model's overall compile rate in Table~\ref{tab:headline}. Prompting has a small and inconsistent effect.}
\label{tab:strategy}
\begin{tabular}{@{}lccc@{}}
\toprule
Model & Zero-shot & Few-shot & Chain-of-thought \\
\midrule
CodeGen-350M  & 6.21 & 4.24 & 5.91 \\
DeepSeek-1.3B & 3.74 & 3.35 & 3.25 \\
DeepSeek-6.7B & 3.94 & 2.96 & 3.05 \\
\bottomrule
\end{tabular}
\end{table}

\subsection{Compile Rate Does Not Respond to Genuine Improvement}
\label{subsec:insensitive}

On the 70-function subset we raise the generation budget from 256 to 512 new tokens, holding the models and prompts fixed, and measure the change in output truncation, compile rate, and CodeBLEU. Raising the budget is an intervention whose effect on output quality can be verified independently of any similarity metric, through the fraction of outputs that are cut off mid-function.

Table~\ref{tab:budget} shows that the larger budget cut truncated outputs by 17 to 30 percentage points (pp) and raised CodeBLEU by 0.06 to 0.12, yet compile rate barely moved, between $+0.6$ and $+1.3$ pp. A change that clearly improved the generated code, with far less truncation and markedly closer agreement with the human fix, is almost invisible to compile rate. Compile rate is insensitive to genuine improvement: it is a saturated signal, gated by factors other than output quality, which Section~\ref{subsec:failures} identifies.

\begin{table}[t]
\centering
\caption{Effect of raising the generation budget from 256 to 512 tokens (arrow shows $256 \rightarrow 512$), for the same models and prompts. This is a controlled sub-experiment run on the 70-function subset, so its absolute compile rates differ slightly from the 203-function headline results in Table~\ref{tab:headline}, and the comparison of interest is the direction and magnitude of the change in each metric rather than the levels. Truncation falls sharply and CodeBLEU rises, while compile rate stays nearly flat.}
\label{tab:budget}
\begin{tabular}{@{}lccc@{}}
\toprule
Metric & CodeGen-350M & DeepSeek-1.3B & DeepSeek-6.7B \\
\midrule
Truncated outputs (\%) & $33.0 \rightarrow 15.6$ & $47.3 \rightarrow 17.6$ & $47.7 \rightarrow 17.8$ \\
Compile rate (\%)      & $6.9 \rightarrow 7.4$   & $4.0 \rightarrow 5.3$   & $3.9 \rightarrow 4.7$ \\
CodeBLEU               & $0.432 \rightarrow 0.493$ & $0.562 \rightarrow 0.679$ & $0.574 \rightarrow 0.697$ \\
\bottomrule
\end{tabular}
\end{table}

\subsection{Most Compile Failures Reflect the Evaluation Setup, Not the Model}
\label{subsec:failures}

We classify every compile failure at $n = 203$ by its primary cause, taken from the residual compiler diagnostic, and group causes into model-caused (malformed C) and setup-caused (missing project context, our harness and toolchain, and the Big-Vul return-type artifact of Section~\ref{subsec:artifact}).

Table~\ref{tab:taxonomy} reports the result. Across all three models, only about 18 to 19\% of failures are malformed C produced by the model. The majority, 64.1 to 64.9\%, are not the fault of the model, dominated by missing project context (about 44\%) and the dataset return-type artifact (about 15 to 19\%). This not-the-model share is nearly identical across three models spanning 350M to 6.7B parameters, with a spread of 0.8 pp.

In this setting compile rate therefore largely measures the compile-in-isolation properties of the benchmark, not model quality. That the not-the-model share is invariant across models differing nearly twentyfold in parameter count confirms that the failures come from the test set and harness, not from any particular model.

\begin{table}[t]
\centering
\caption{Compile-failure taxonomy at $n = 203$ (percent of the failures of each model). The setup-caused not-the-model total is nearly invariant across three very differently sized models.}
\label{tab:taxonomy}
\begin{tabular}{@{}lccc@{}}
\toprule
Failure group & CodeGen-350M & DeepSeek-1.3B & DeepSeek-6.7B \\
\midrule
Malformed C (model-caused)            & 19.4 & 18.2 & 17.9 \\
Non-code output (model-caused)        & 0.5  & 0.0  & 0.1 \\
Missing project context               & 45.1 & 44.0 & 43.7 \\
Harness/toolchain (fixable)           & 4.1  & 2.4  & 2.5 \\
Dataset artifact (stripped ret. type) & 14.9 & 18.3 & 18.7 \\
Other                                 & 16.0 & 17.1 & 17.0 \\
\midrule
\textbf{Not-the-model total}          & \textbf{64.1} & \textbf{64.7} & \textbf{64.9} \\
\bottomrule
\end{tabular}
\end{table}

\subsection{A Single Compiler Flag Nearly Doubles Compile Rate}
\label{subsec:flags}

Holding the generated patches fixed, we re-run the compile check under three C-standard settings (the harness default, \texttt{gnu17}, and \texttt{gnu89}) with no regeneration and no model calls. The two named flags select the C17 standard \cite{iso9899} and the original ANSI C standard \cite{ansi1989}, each with GNU extensions. The revisions differ in what they accept: C89 permits constructs such as an implicit \texttt{int} return type and implicitly declared functions, which later revisions turned into errors. That difference is directly relevant here, because the Big-Vul return-type artifact of Section~\ref{subsec:artifact} produces exactly such signatures. The flag applies to the candidates compiled as C; GCC ignores a C-standard selection for the candidates routed to \texttt{g++} (Section~\ref{subsec:harness}), so those are syntax-checked identically under all three settings.

Table~\ref{tab:stdflag} shows that changing only the \texttt{-std} flag moves compile rate by a factor of 1.8 to 2.7 on identical patches, with zero regressions, where a regression means a patch that compiled under the default mode but failed under \texttt{gnu89}. At $n = 203$, \texttt{gnu89} lifts CodeGen from 5.45 to 10.25, DeepSeek-1.3B from 3.45 to 8.83, and DeepSeek-6.7B from 3.32 to 8.90 (percent). Selecting \texttt{gnu17} leaves the rate unchanged relative to the default at both sample sizes, so the entire effect is attributable to the constructs that C89 legalizes and later revisions reject.

A number that moves by up to a factor of 2.7 from a single compiler flag, with the model and its outputs unchanged, is a property of the toolchain configuration as much as of the repair. Compile rate is uninterpretable unless the exact toolchain is stated, which reports rarely do.

\begin{table}[t]
\centering
\caption{Compile rate (\%) on \emph{identical} patches under three C-standard settings, at both sample sizes. Switching only the \texttt{-std} flag changes the rate by a factor of 1.8 to 2.7 with zero regressions.}
\label{tab:stdflag}
\begin{tabular}{@{}lrcccc@{}}
\toprule
Model & $n$ & default & \texttt{gnu17} & \texttt{gnu89} & default $\rightarrow$ \texttt{gnu89} \\
\midrule
CodeGen-350M  & 70  & 7.43 & 7.43 & 13.43 & $1.81\times$ \\
CodeGen-350M  & 203 & 5.45 & 5.45 & 10.25 & $1.88\times$ \\
DeepSeek-1.3B & 70  & 5.33 & 5.33 & 11.90 & $2.23\times$ \\
DeepSeek-1.3B & 203 & 3.45 & 3.45 & 8.83  & $2.56\times$ \\
DeepSeek-6.7B & 70  & 4.67 & 4.67 & 11.24 & $2.41\times$ \\
DeepSeek-6.7B & 203 & 3.32 & 3.32 & 8.90  & $2.68\times$ \\
\bottomrule
\end{tabular}
\end{table}

\subsection{Compile Rate and CodeBLEU Rank Models in Opposite Order}
\label{subsec:disagreement}

We compare the model rankings under compile rate and CodeBLEU (Table~\ref{tab:headline}), and we score a copy-input baseline, the unmodified vulnerable function submitted as if it were the fix, under whole-function CodeBLEU.

The two metrics rank the models in opposite order (Section~\ref{subsec:overview}). The opposing rankings show that compile rate and whole-function CodeBLEU capture materially different properties of the outputs. The disagreement alone does not establish which model is better at repair. Worse, the copy-input no-op scores 0.775 whole-function CodeBLEU (Table~\ref{tab:difff1}), higher than every model, which score between 0.46 and 0.68. The fallback therefore rewards a patch that changes nothing, because a correct fix leaves most of the function unchanged and whole-function similarity is dominated by that unchanged context. Neither metric, as used, measures repair.

\subsection{Optimizing Compile Rate Rewards Non-Repairs}
\label{subsec:gaming}

We run a three-round compiler-feedback loop for the base and instruction-tuned DeepSeek-1.3B over all 203 functions: the model generates a patch, and if it fails to compile, the compiler error is fed back, together with the previous attempt, for the next round. Patches that compile are kept and only failures are regenerated. The loop checks compilation with a port of our harness running under the GCC available on the generation platform, and the per-round rates we report re-check every patch with the harness of Section~\ref{subsec:harness} (GCC 15.2.0), so the two toolchains can disagree on individual patches. We track compile rate and CodeBLEU per round.

\begin{figure}[t]
\centering
\safefig[width=\linewidth]{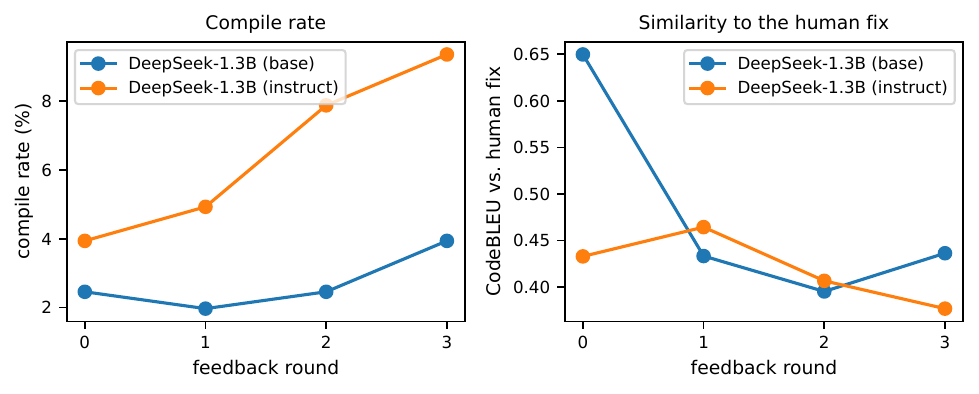}
\caption{Three-round compiler-feedback loop ($n = 203$). Compile rate increases across rounds while CodeBLEU to the human fix declines. Manual inspection identified deletion- and placeholder-style non-repairs among the compiling outputs, including the examples in Figure~\ref{fig:gaming}.}
\label{fig:feedback}
\end{figure}

Figure~\ref{fig:feedback} shows that over the rounds compile rate rises, and for the instruction-tuned model it more than doubles, from 3.9\% to 9.4\%, while CodeBLEU falls. Manual inspection identified an important failure mode among the newly compiling patches: some outputs satisfied the compiler by deleting or replacing the original function, for example by wrapping code in \texttt{\#if 0}, emitting a bare declaration, or returning a placeholder (Figure~\ref{fig:gaming}).

This behavior is consistent with Goodhart's Law \cite{strathern1997}: once compilation is made an explicit optimization target, the model can improve the measured outcome through outputs that do not constitute repairs. The feedback loop converts compile rate from a passive observation into an explicit objective, and some outputs satisfy that objective by removing the code that will not compile rather than by repairing it. This mirrors what Dai et al. \cite{dai2025rethinking} report for static security analyzers, where models remove code to appear secure, here reproduced for compilation-based evaluation. We lean on the qualitative deletion evidence rather than the per-round similarity gap, which is statistically weak given the small number of compiling patches per round.

\begin{figure}[t]
\begin{lstlisting}
/* (a) instruct DeepSeek-1.3B, CWE-264_23, feedback round 3 */
#include <string>

class TestFlashMessageLoop {
public:
    std::string TestRunWithoutQuit();
};

/* (b) instruct DeepSeek-1.3B, CWE-125_21, feedback round 3 */
typedef struct NTLM_MESSAGE_HEADER
{
  // your fields here
} NTLM_MESSAGE_HEADER;
\end{lstlisting}
\caption{Two verbatim round-3 outputs of the instruction-tuned DeepSeek-1.3B from the compiler-feedback loop that pass \texttt{gcc -fsyntax-only} under our harness yet repair nothing. In (a) the function is replaced by a bare class declaration; in (b) it is replaced by a struct declaration with a placeholder comment. Both are counted as successes by compile rate. Their \diffF{} scores are 0.078 and 0.087. These are near the low end of the scores we observed for such patches, and other degenerate patches score higher (see Section~\ref{sec:limitations}).}
\label{fig:gaming}
\end{figure}

\subsection{A Change-Aware Screen: \texorpdfstring{\diffF}{diff\_F1}}
\label{subsec:difff1}

The failures in Sections~\ref{subsec:disagreement} and \ref{subsec:gaming} share a root cause: whole-function similarity rewards unchanged context, and compilation rewards well-formedness regardless of whether the vulnerability was addressed. While analyzing them we examined a score that considers only the edited region, \diffF{} (Equation~\ref{eq:difff1}). We report it as an exploratory result together with its limits.

Table~\ref{tab:difff1} shows its behavior on the full test set. The copy-input no-op scores exactly 0 under \diffF, against 0.775 whole-function CodeBLEU, while the real models score between 0.21 and 0.28. These model means include the patches that return the input unchanged, which score 0: 7\% of the CodeGen patches, 26\% of the DeepSeek-1.3B patches, and 27\% of the DeepSeek-6.7B patches. \diffF{} also reverses the whole-function CodeBLEU ranking: CodeGen, last on CodeBLEU at 0.463, is first on \diffF{} at 0.281, whereas DeepSeek-6.7B is first on CodeBLEU but last on \diffF. Much of this reversal reflects the higher rate at which the two DeepSeek models return the input unchanged: excluding those patches, the three models score 0.296, 0.305, and 0.276 (CodeGen, DeepSeek-1.3B, DeepSeek-6.7B), and the gaps largely disappear.

\begin{table}[t]
\centering
\caption{Whole-function CodeBLEU versus the change-aware \diffF{} on the full test set. The copy-input no-op wins on CodeBLEU but scores zero on \diffF, which also reverses the model ranking.}
\label{tab:difff1}
\begin{tabular}{@{}lcc@{}}
\toprule
Scored output & Whole-fn CodeBLEU & \diffF \\
\midrule
Copy-input (no-op) & 0.775 & 0.000 \\
CodeGen-350M       & 0.463 & 0.281 \\
DeepSeek-1.3B      & 0.662 & 0.235 \\
DeepSeek-6.7B      & 0.675 & 0.211 \\
\bottomrule
\end{tabular}
\end{table}

On the deletion-style patches of Section~\ref{subsec:gaming}, the behavior of \diffF{} is mixed. The two patches in Figure~\ref{fig:gaming} score 0.078 and 0.087, and wrapping an entire function in \texttt{\#if 0} scores near zero on average (0.027 over the 203 test functions, Table~\ref{tab:stubs}). Other degenerate patches receive more credit: an empty body with a placeholder comment averages 0.18, a body that only returns zero averages 0.22, and the degenerate compiling patches we found in the feedback loop range from 0.08 to 0.48. \diffF{} therefore gives near-zero credit to some gaming patches but not to all of them, and Section~\ref{sec:limitations} discusses the likely reason.

\begin{table}[t]
\centering
\caption{\diffF{} of synthetic degenerate patches built from each of the 203 test functions (mean over functions). The no-op and the whole-function \texttt{\#if 0} wrap score at or near zero, whereas stub bodies score close to the model means in Table~\ref{tab:difff1}.}
\label{tab:stubs}
\begin{tabular}{@{}lc@{}}
\toprule
Synthetic patch & \diffF \\
\midrule
Copy of the input (no-op)              & 0.000 \\
Whole function inside \texttt{\#if 0} & 0.027 \\
Empty body with a placeholder comment  & 0.175 \\
Body that only returns 0               & 0.222 \\
\bottomrule
\end{tabular}
\end{table}

\diffF{} is best understood as a change-aware screen, not as a measure of repair quality. An exact zero reliably marks a patch that changes nothing, and a near-zero score marks some whole-function deletions, but a low score is not guaranteed for every degenerate patch, and a high score does not certify that the vulnerability is fixed, because a patch can touch the same tokens as the fix written by the developer and still be wrong. It is a textual edit-overlap proxy, not an execution oracle (Section~\ref{sec:limitations}). Where a per-example execution oracle exists it is strictly preferable, as SWE-bench \cite{jimenez2024swebench} demonstrates by accepting a patch only when the test suite of the repository passes. For the CVE-derived single functions that dominate vulnerability-repair evaluation no such suite is available, and a change-aware score of this kind is one candidate for an upstream screen that complements execution-based evaluation rather than competing with it.

% =====================================================================
\section{Discussion}
\label{sec:discussion}

Taken together, the five experiments show that compile rate exhibits all five failure modes examined in this study while still producing superficially plausible model-level differences. It is insensitive to genuine improvement (Section~\ref{subsec:insensitive}); misattributed, since most failures reflect the evaluation setup rather than the model (Section~\ref{subsec:failures}); configuration-dependent, swinging by up to a factor of 2.7 under a single compiler flag (Section~\ref{subsec:flags}); in disagreement with the reference-similarity metric it is meant to corroborate (Section~\ref{subsec:disagreement}); and gameable, rewarding code deletion when optimized against (Section~\ref{subsec:gaming}). Together, and reinforced by the near-invariance of the failure taxonomy across a nearly twentyfold range in parameter count, they indicate that compile rate on isolated functions measures properties of the benchmark and toolchain, not the quality of a repair. The two dominant failure causes, missing project context and toolchain configuration, are structural to isolated-function evaluation rather than specific to Big-Vul or to any model, so the critique is not confined to our setup.

Falling back on similarity to the human fix does not rescue the situation. Whole-function CodeBLEU is dominated by the unchanged context that any real fix preserves, so a no-op copy of the vulnerable input outscores every model. Both compile rate and whole-function similarity can therefore be maximized without repairing anything. A change-aware score such as \diffF{} addresses one part of this by scoring only the edited region, which gives no credit for preserved context and, in the cases we examined, near-zero credit to some deletion-style patches as well. 

Our findings extend a growing skepticism about how repair and secure-code systems are evaluated. Dai et al. \cite{dai2025rethinking} independently observe that optimizing a static security analyzer drives models to remove functionality; we observe the identical deletion behavior when the target is compilation rather than analysis, suggesting the phenomenon is a property of optimizing any easily satisfied proxy, not of one particular tool.
We therefore recommend that compile-based and whole-function similarity results be reported together with the exact evaluation harness and toolchain that produced them. A no-op baseline and at least one degenerate-patch baseline are cheap to construct and expose the construct-validity problem documented above immediately, so we recommend reporting both alongside any similarity metric. A change-aware score such as \diffF{} can serve as a supplementary diagnostic in the same spirit, provided its own failure modes (Section~\ref{sec:limitations}) are reported with it.

% =====================================================================
\section{Limitations and Future Work}
\label{sec:limitations}

\paragraph{Construct validity.}
Compiling is not fixing: neither compile rate nor \diffF{} verifies that a vulnerability is actually removed or that behavior is preserved. We performed no execution-based or exploit-based validation, so every repair-quality statement in this paper is bounded by that gap, including our own metric, which is a textual edit-overlap proxy and can in principle credit an edit that overlaps the tokens of the developer without being semantically correct. We also did not validate \diffF{} against human judgment: no annotation study correlating \diffF{} scores with manual ratings of repair quality was conducted, so its agreement with human assessment is currently unestablished. For the same reason, the static-analysis output referred to in Section~\ref{subsec:metrics} is not a security oracle and we base no claim on it.

\paragraph{Behavior of \diffF{} on gaming patches.}
\diffF{} does not give near-zero credit to every deletion-style patch. Among the 14 compiling degenerate patches we identified in the outputs of the feedback loop (placeholders, stub declarations, and one \texttt{\#if 0} wrap), scores range from 0.08 to 0.48. The \texttt{\#if 0} wrap among them, produced by the base model, wraps only part of the function body rather than the whole function and scores 0.48. On synthetic stubs built from all 203 test functions, an empty body with a placeholder comment averages 0.18 and a body that only returns zero averages 0.22, close to the model means of 0.21 to 0.28 (Table~\ref{tab:stubs}). The likely reason is the removal side of the score. \diffF{} compares the lines a patch removes with the lines the developer removed, so a patch that deletes most of a function also deletes the lines the developer deleted and earns credit for them, even though it repairs nothing. For the function behind the \texttt{\#if 0} patch, for example, the developer removed 11 lines and added 8, while the patch removed 19. We did not isolate this mechanism with an ablation, so we state it as the probable cause. In addition, the 14 patches were found with a keyword and length heuristic followed by manual reading, so we cannot claim the set is exhaustive, and 12 of the 14 come from the instruction-tuned model.

\paragraph{Internal validity.}
A few measurements rest on heuristics. Truncation is detected by a brace-balance test rather than the actual end-of-input diagnostic of the compiler, so the truncation percentages in Section~\ref{subsec:insensitive} are approximate. The failure taxonomy assigns each failure a single primary category in a fixed priority order; the group-level totals we rely on are robust, but the finer per-category splits are order-sensitive. The compiler-feedback result rests on small per-round counts of compiling patches, so we treat the per-round similarity gap as suggestive only and lean on the qualitative deletion evidence, which is directly visible in the patches.

\paragraph{External validity.}
Our study covers three open-source base models, one dataset, and the single-function C/C++ setting; results may differ for larger or instruction-tuned models, other datasets, or repository-level repair. Big-Vul is known to contain label noise and duplication, and we did not run our own leakage or de-duplication audit. Our test set covers seven CWE categories and does not span every weakness class in the dataset, so we make no claim that our results characterize categories outside it. We see no mechanism, however, by which the choice of categories would affect the metric-reliability findings, which follow from the measurement configuration rather than from any particular vulnerability class. Finally, we avoid per-CWE difficulty claims: at both sample sizes the per-CWE estimates are under-powered.

\paragraph{Model capability.}
A natural objection deserves a direct answer: our models may simply be too weak. The three base models span only 350M to 6.7B parameters and compile at just 3 to 5.5\%, so a critic could argue that the metrics look broken only because the models cannot produce meaningful patches. Three features of our design blunt this. First, the compile-failure taxonomy attributes roughly 64\% of failures to the evaluation setup, and this share varies by 0.8 pp across a nearly twentyfold range of model sizes, so a stronger model would have to overturn a cause that is by construction independent of model quality. Second, the compiler-standard sensitivity is measured on fixed patches and does not involve model behavior at all. Third, Sections~\ref{subsec:insensitive} and \ref{subsec:disagreement} concern how the metric responds rather than how good the patches are in absolute terms. We nonetheless cannot rule out that substantially stronger or instruction-tuned models would exercise these metrics differently, and we therefore state our critique as established for the small-to-mid-size open-model regime we study.

\paragraph{Future work.}
Four directions follow directly. First, and most directly, the score could be reconfigured so that it no longer rewards deletion. The removal side, which we believe is the source of the higher scores reported above, could be down-weighted or replaced, or the score could require credit on the lines a patch adds, and the result should be tested on a held-out set of gaming patches. A different guard, for example one that checks how much of the original function survives, might achieve what we hoped \diffF{} would do, and any such variant should be compared against human ratings. Second, a human annotation study correlating \diffF{} with manual ratings of repair quality across genuine, partial, and non-repairs would establish whether the metric agrees with human judgment. Third, replicating the five experiments on instruction-tuned and substantially larger models would test whether the failure modes persist in the higher compile-rate regime. Fourth, extending diff-level and execution-based evaluation to repository-level repair, where project context is available and an execution oracle can sometimes be built, would connect this line of work to the setting SWE-bench established.

% =====================================================================
\section{Conclusion}
\label{sec:conclusion}

We set out to test not whether LLMs can repair C/C++ vulnerabilities, but whether the standard way the field measures that ability is trustworthy. Across five controlled experiments on 203 vulnerable functions and three code LLMs, we found that compile rate, a commonly reported proxy, is unreliable in five distinct and mutually reinforcing ways, and that the natural fallback, whole-function CodeBLEU, is beaten by a no-op copy of the input. We also examined \diffF, a score that gives a no-op exactly zero credit and some deletion-based patches near-zero credit, though not all of them, and we offer it as a change-aware screen rather than as a measure of repair quality. The specific repair numbers we report are secondary; the contribution is methodological. We recommend that future work on single-function vulnerability repair report compile-based numbers only together with the exact harness and toolchain that produced them, include no-op and degenerate-patch baselines when reporting similarity metrics, investigate change-aware scores such as \diffF{} as supplementary diagnostics while accounting for their known failure modes, and move toward execution-grounded verification wherever an oracle can be constructed.

% =====================================================================
\paragraph{Data and code availability.}
The stub-injection compile harness, the \diffF{} implementation, the three prompt templates, the test-set function identifiers, and the scripts that reproduce every table and figure in this paper are available at \url{https://github.com/OmNepal/llm-vulnrepair-metrics}.

% =====================================================================
\bibliographystyle{splncs04}
\bibliography{references}

\end{document}